\documentclass[11pt,aps,pra,onecolumn,citeautoscript,superscriptaddress,a4paper]{revtex4-1}

\usepackage{amsmath}
\usepackage{amssymb}
\usepackage{amsfonts}
\usepackage{mathtools}
\usepackage[b]{esvect}
\usepackage{pst-node}
\usepackage[utf8]{inputenc}
\usepackage[english]{babel}
\usepackage{graphicx}
\usepackage{subcaption}
\usepackage{xcolor}
\usepackage{braket}
\usepackage[colorlinks=true,allcolors=black]{hyperref} 
\usepackage[left=2.0cm,right=2.0cm,top=2.3cm,bottom=2.0cm]{geometry}
\usepackage{bbold}

\usepackage{array} 
\newcolumntype{C}[1]{>{\centering\let\newline\\\arraybackslash\hspace{0pt}}m{#1}}
\newcolumntype{N}{@{}m{0pt}@{}}
\usepackage{breqn}
\usepackage{caption}
\usepackage[toc,page]{appendix}
       
\begin{document}

\author{Seth Lloyd}
\affiliation{Department of Mechanical Engineering$,$ Massachusetts Institute of Technology$,$ Cambridge$,$ Massachusetts 02139$,$ USA}
\affiliation{Department of Physics$,$ Massachusetts Institute of Technology$,$ Cambridge$,$ Massachusetts 02139$,$ USA}
\author{Reevu Maity}
\affiliation{Clarendon Laboratory$,$ University of Oxford$,$ Parks Road$,$ Oxford$,$ OX1 3PU$,$ UK} 

\title{\bf Efficient implementation of unitary transformations}

\begin{abstract}
 Quantum computation and quantum control operate by building unitary transformations out of sequences of elementary quantum logic operations or applications of control fields. This paper puts upper bounds on the minimum time required to implement a desired unitary transformation on a $d$-dimensional Hilbert space when applying sequences of Hamiltonian transformations.  We show that strategies of building up a desired unitary out of non-infinitesimal and infinitesimal unitaries, or equivalently, using power and band limited controls, can yield the best possible scaling in time $O(d^2)$.    
\end{abstract}

\maketitle

\bigskip

\section{Introduction}
One of the main applications of quantum computation and quantum control in real world problems is to simulate the dynamics of physical systems
\cite{Feynman,Lloyd1}. In quantum computation, one can build any desired unitary operation from a sequence of quantum gates, $U = U_N...U_1$ where $U_j$ represents an elementary quantum logic gate. In practice, logic gates that we apply are of the form $U_j = e^{-i H_j t_j}$ where $H_j$ is a k-local Hamiltonian acting on a physical system for time $t_j$. A closely related problem appears in the context of continuous time quantum control where one applies a time dependent Hamiltonian of the form $H(t) = \sum_j g_j(t) H_j$. Here $g_j(t)$ corresponds to a time-dependent control field. For constructing an arbitrary unitary in both quantum computation and quantum control, a necessary and sufficient condition is that the algebra generated by the Hamiltonians $\{H_j\}$ via commutation should be complete in $u(d)$. This is because in general, we are not concerned about an overall global phase. In this work, we restrict our attention to traceless Hamiltonians in $su(d)$. 

A major challenge in the fields of quantum computation and quantum control is to construct arbitrary unitary transformations efficiently. It is well known that one can simulate local Hamiltonians in time that is poly-logarithmic in the dimension of the physical system 
\cite{Lloyd1,Aharonov,Childs2,Berry1,Childs3,Wiebe1,Childs4,Wiebe2,Poulin,Berry3,Childs5}. A significant amount of literature till date has been dedicated to the study of optimal construction of unitary transformations in quantum computation and quantum control \cite{Shende,Mottonen,Hanneke,Georgescu,Pechen,Wei,Zhang}. Recent works have significantly improved the dependence of gate complexity on the precision for sparse and low-rank Hamiltonian dynamics simulation \cite{Berry4,Berry2,Low1,Low2,Haah,Low3,Rebentrost1,Rebentrost2,Childs6,Childs7}. There has also been a spate of activities in the fields of trapped ions and superconducting quantum computing to advance the implementation of two-qubit logic gates for generating quantum entanglement and fault-tolerant quantum computation \cite{Schafer,Gambetta}. The discrete version for approximating arbitrary unitaries in $SU(d)$ is covered by the Solovay-Kitaev theorem \cite{Kitaev,Nielsen,Dawson} and its corresponding inverse-free versions \citep{Sardharwalla,Bouland}. We investigate the continuous version of this problem for building arbitrary unitary matrices using sequences of non-commuting Hamiltonian operations. 

In this work, we address the question of the optimal time required to implement an arbitrary unitary operator in $d$ dimensions. We further ask whether one can find the correct sequence of logic gates or control fields that can be applied in order to generate the target unitary matrix. We know that $d^2-1$ parameters are required to specify an arbitrary unitary matrix in $SU(d)$. We present two approaches, namely the non-infinitesimal and infinitesimal unitary methods for simulating a desired unitary transformation in $SU(d)$. In the infinitesimal approach, we build up transformations in the vicinity of the identity using nested commutation relations. In the non-infinitesimal approach, we move away from the identity, perform a set of transformations and then return. The time complexity of the non-infinitesimal method scales as $O(d^2)$ while the scaling with dimension is $O(d^2 \log d)$ for the infinitesimal approach. Our first result is to show that one can generically construct any desired unitary matrix in the neighbourhood of identity in time that scales as $O(d^2)$ by appropriately choosing a sequence of parameters in $\vec{t}, \vec{\tau}$. This can be achieved either by a direct method when there is an explicit representation of the unitary or by a process of gradient descent when the unitary is described by a training set of input and output pairs. 

In quantum control, a remarkable result by Rabitz {\it et al.} \cite{Rabitz1,Rabitz2} 
demonstrates that when the control fields $g_j(t)$ are neither power nor
band limited, the optimal control sequence for constructing a desired unitary can be achieved by the method of gradient descent. In practice, however, control fields are power and band limited. Our second result is to demonstrate that even with power and band limited controls, one can construct any desired unitary near identity in time of $O(d^2)$ by gradient descent. We emphasize that the primary differences between the infinitesimal and non-infinitesimal results is that the infinitesimal technique takes longer time to reach a particular unitary and the process of finding a viable path to construct the unitary using gradient descent involves a search through multiple saddle points. 

We consider the simplest case when there are two non-commuting traceless Hamiltonians $A$ and $B$. Our results generalize in straightforward fashion to the case of three or more Hamiltonians. We construct unitaries of the form following \cite{Suzuki1, Suzuki2}
\begin{equation} \label{eq:1}
U(\vec{t},\vec{\tau}) = e^{-i B \tau_N} e^{-i A t_N} \ldots e^{-i B \tau_1} e^{-i A t_1} .
\end{equation}
Our first assumption is that the operators $A$ and $B$ in $\mathcal{H}_d$ are bounded such that $\| A \|_1$=1, $\|B\|_1 = 1$. The second assumption is that $A$ and $B$ generate the entire Lie algebra of $su(d)$. For example, $A$ and $B$ could be random matrices with elements selected from a Gaussian ensemble and appropriately scaled such that they have unit 1-norm. We further assume that both $\pm A$ and $\pm B$ can be implemented so that $\{t_i,\tau_j\}$ can have positive or negative signs depending on the target unitary that we aim to construct. In the rest of the paper, we will consider the parameters $\{t_i,\tau_j\}$ to be positive.

The parameters $\{t_i,\tau_j\}$ can be large in the quantum logic gate construction of unitary operators. In continuous time quantum optimal control, the Hamiltonian dynamics governed by a time dependent Hamiltonian $g(t) A + h(t) B$ can be implemented in the small time, large $N$ limit of (\ref{eq:1}) by the familiar process of Trotterization \cite{Lloyd1,Nielsen,Wiebe1,Lloyd2}. In this case, we represent the time dependent dynamics of a sequence of infinitesimal transformations.

\section{General Approach}
\begin{figure}[t]
\begin{subfigure}{0.45\linewidth}
\includegraphics[width=\linewidth,height=5cm]{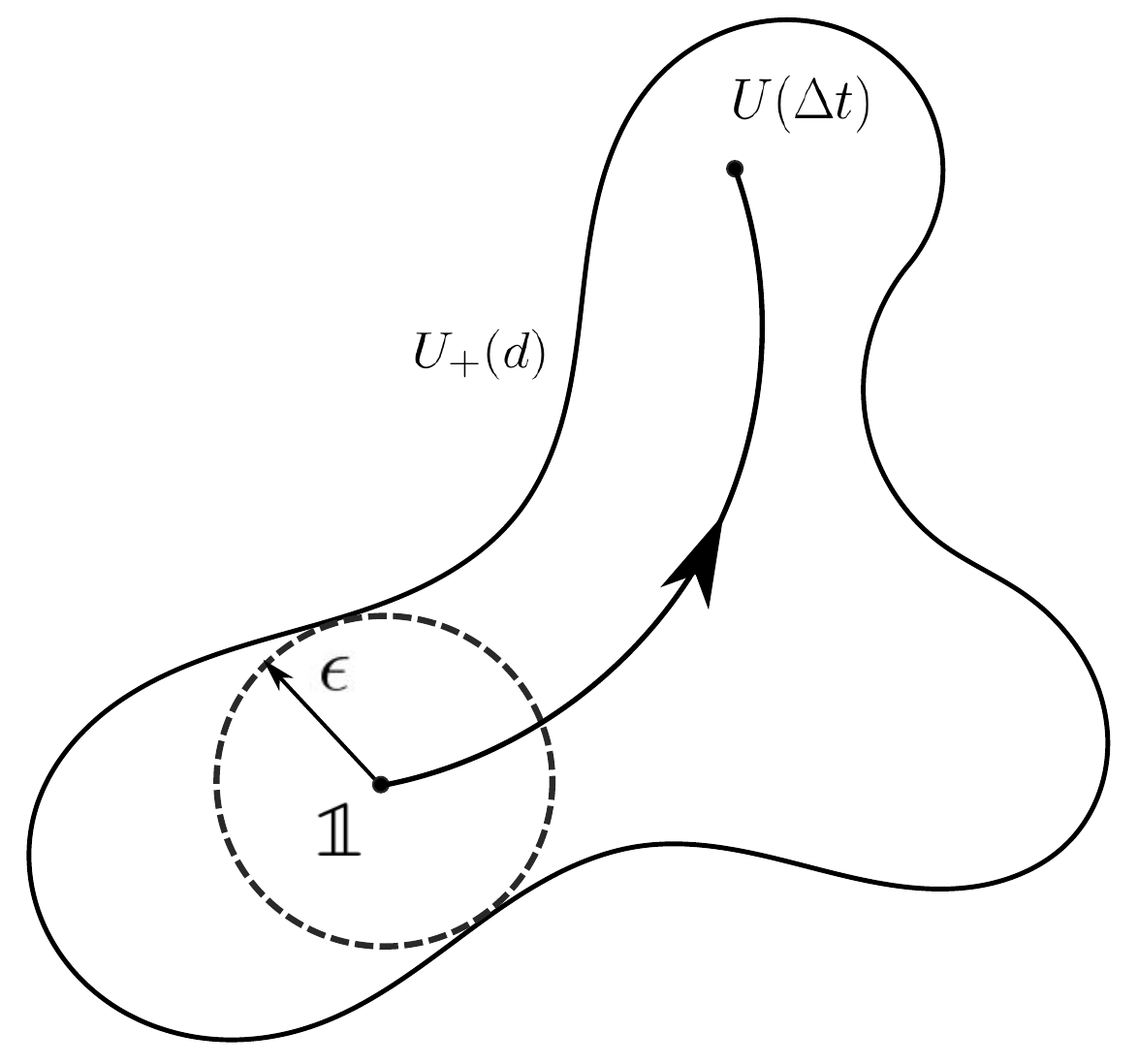}  
\caption{}
\end{subfigure}
\begin{subfigure}{0.45\linewidth}
\includegraphics[width=\linewidth,,height=5cm]{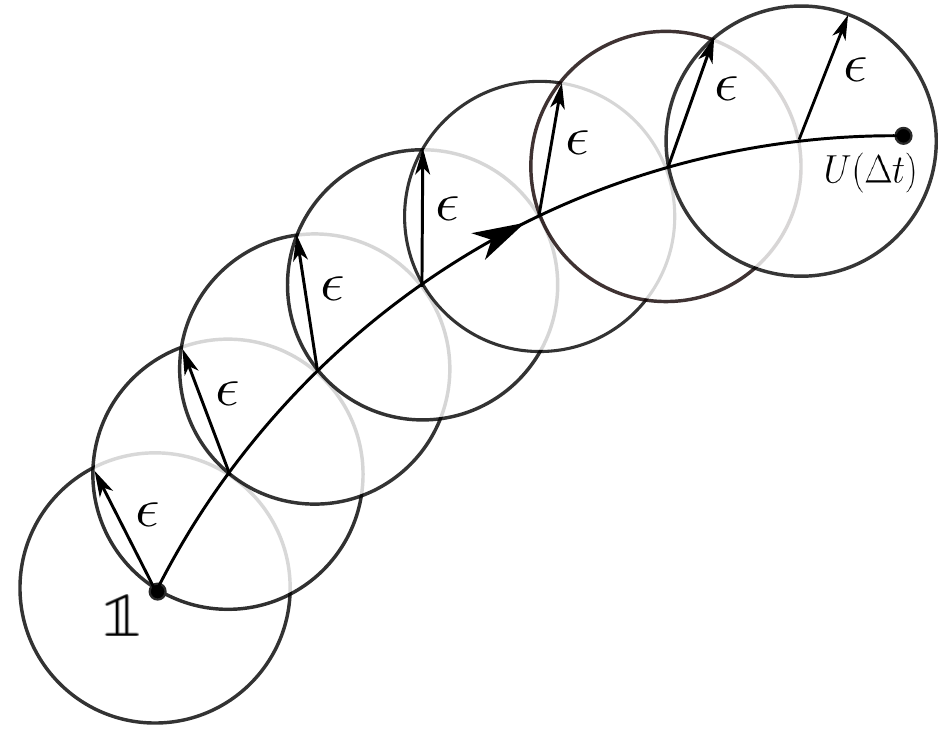}
\caption{}
\end{subfigure}%
\caption{a) Largest $\epsilon$-ball around $\mathcal{I}$ in the space of time-polynomial unitaries $U_+(d)$ reachable in time $\Delta t$. b) $\epsilon$-balls covering a path of finite length $L$ in $U_+(d)$ from initial unitary $\mathcal{I}$ to final unitary $U(\Delta t)$.}
\label{fig1}
\end{figure}

In this section, we provide an outline of our constructive approach for both the non-infinitesimal and infinitesimal methods. We show that when $N=O(d^2)$ there exists an $\epsilon$-ball in the vicinity of the identity operator and $\epsilon > 0$ such that one can construct any desired unitary operator given by $e^{-i H t}$ where $\| H \|_1$=1 and $t \leq \epsilon$, by suitable choice of $\{t_i,\tau_j\}$ in (\ref{eq:1}). The size of $\epsilon$ depends on the particular control Hamiltonians $A, B$ that we can implement. The essential point here is that the radius of the $\epsilon$-ball should be strictly bounded away from zero. In practice, because of the finite precision achievable in $\{t_i,\tau_j\}$, we can only reach the desired unitary approximately. The size of $\epsilon$ and the effects of such finite precision will be addressed below.   

In other words we show that it is possible to advance a non-zero distance along any direction in the algebra of Hamiltonians in $su(d)$. The actual unitary that we aim to implement is given by $U = e^{-i H t}$, $|t| \leq \pi$. To perform this, we first find a sequence of gates or control parameters to realize $e^{-i H \epsilon}$ below. We demonstrate how to construct $e^{-i H \epsilon}$ when there is an explicit representation of $U$. Let $U_+(d)$ be the set of time-polynomial reachable unitaries in $SU(d)$. For elements belonging to the subgroup $U_{+}(d)$, there exists a path of finite length that connects the initial and final unitaries in finite time \cite{Lloyd4}. We can then simply repeat the sequence $t/\epsilon$ times to implement $e^{-i H t}$. Figure \ref{fig1} illustrates our general method for both the non-infinitesimal and infinitesimal approaches to construct a desired unitary operation in the $SU(d)$ manifold.

In the non-infinitesimal case, as will be seen, we typically require $2 d^2$ terms in (\ref{eq:1}) to reach any unitary transformation within $\epsilon = O(1)$ of the identity. In the infinitesimal case, we can build unitaries within $O(\epsilon/ \log d)$ of the identity with $N=O(d^2)$ terms. Consequently, we require $N = O(d^2/ \epsilon)$ steps to construct $U$ while in the infinitesimal case, we need $N = O(d^2 \log d/ \epsilon)$. Note that the non-infinitesimal method gives us the best possible scaling as a function of $d$.

\section{Construction: Non-Infinitesimal Case}
\begin{figure}[t]
\includegraphics[width=0.45\linewidth,height=5cm]{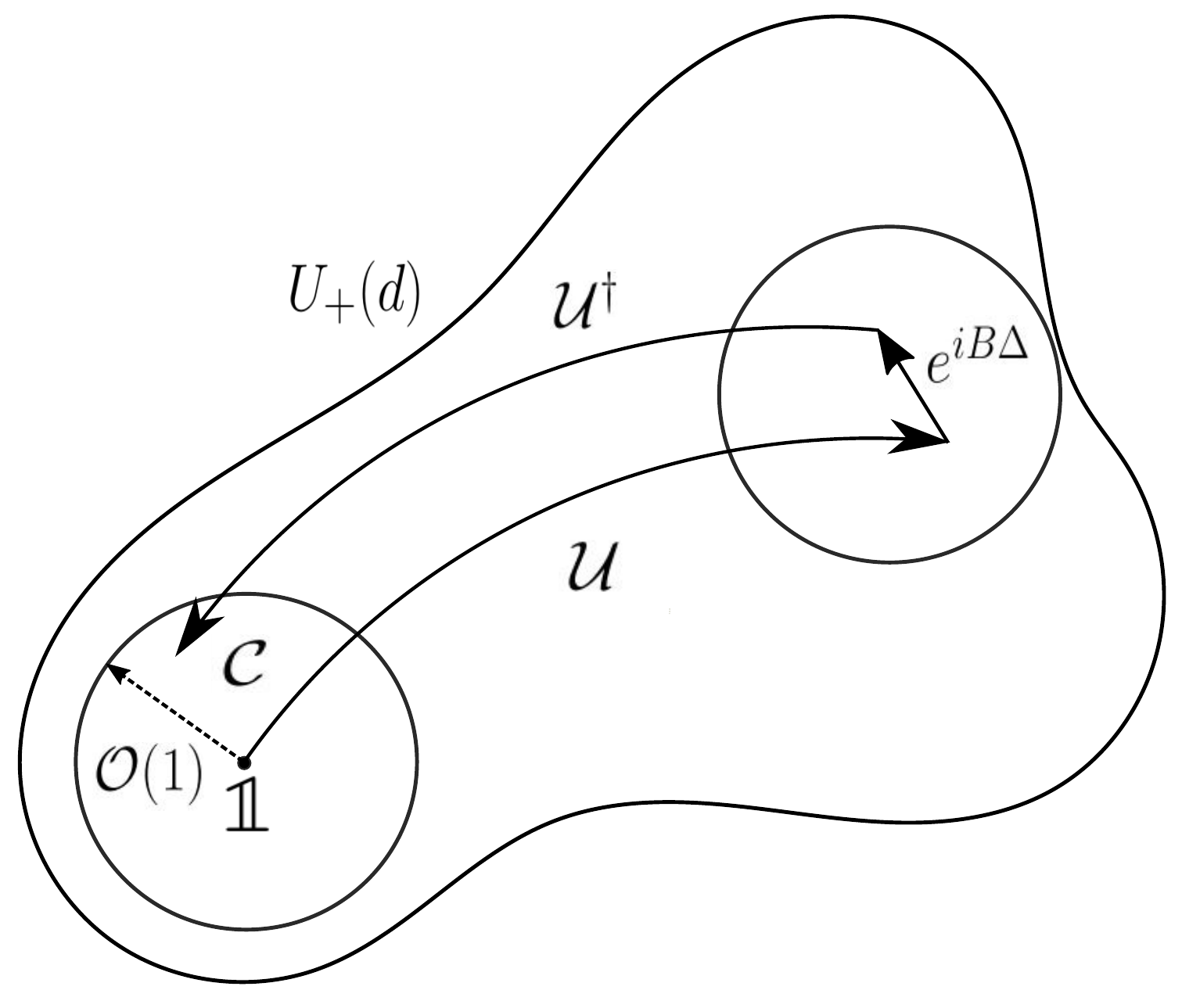}  
\caption{Non-infinitesimal case: Any unitary can be reached within an $\epsilon$-ball around identity where $\epsilon = \mathcal{O}(1)$. First, we prefer a generic $U(\vec{t},\vec{\tau})$. The gradient of $U(\vec{t},\vec{\tau})$ is typically non-zero in all directions. Consequently, we can reach an $\epsilon$ ball of size $\mathcal{O}(1)$ around $U(\vec{t},\vec{\tau})$. Mapping the ball back to the origin $\mathbb{1}$ allows us to attain any point within an $\mathcal{O}(1)$ $\epsilon$-ball of the origin.} 
\label{fig2}
\end{figure}

First, we elucidate the non-infinitesimal technique of constructing any unitary in $SU(d)$. The key point of this argument is, generically $U(\vec{t},\vec{\tau})$ explores a $2 N$ dimensional manifold in the space of all unitaries for $2 N \leq d^2$. When $2 N \geq d^2$, there exists an $\epsilon$-ball of reachable unitaries around a typical point in the manifold. Random selection of $\vec{t}, \vec{\tau}$ can create any unitary within $\epsilon$ of $U(\vec{t},\vec{\tau})$ using $N = O(d^2)$ steps. If an accidental choice of $\vec{t}, \vec{\tau}$ gives a lower dimensional manifold of transformations in the vicinity of $U(\vec{t},\vec{\tau})$, we can discard this selection and choose again. An ideal selection would be $\vec{t},\vec{\tau}$ that maximizes the radius of the $\epsilon$-ball of attainable transformations. Finding this optimal choice is a difficult problem. However, a random choice is adequate for our purpose. Now map this reachable $\epsilon$-ball back to the origin using the inverse transformation $U^{\dagger}(\vec{t},\vec{\tau})=U(-\vec{t},-\vec{\tau}) = e^{i A t_1} e^{i B \tau_1}...e^{i A t_N} e^{i B \tau_N} $. The non-infinitesimal method enables us to realize any transformation $\mathcal{C}$ given by
\begin{subequations} \label{eq:2}
\begin{align} 
\mathcal{C} &= \Big( e^{i A t_1} e^{i B \tau_1} \ldots e^{i A t_N} e^{i B \tau_N} \Big) \Big(e^{-i B \tau_N} e^{-i A t_N} \ldots e^{-i B (\tau_k+\Delta)} e^{-i A t_k} \ldots e^{-i B \tau_1} e^{-i A t_1} \Big) \label{eq:2a}
 \\
&=  \underbrace{e^{i A t_1} e^{i B \tau_1} \ldots e^{i A t_k}}_{\mathcal{U}}.e^{-i B \Delta}.\underbrace{e^{-i A t_k} \ldots e^{-i B \tau_1} e^{-i A t_1}}_{\mathcal{U}^{\dagger}}  \label{eq:2b}
 \\
 &=  e^{-i \tilde{B} \Delta} \hspace{2mm} \mbox{where} \hspace{2mm} \tilde{B} = \mathcal{U} B \mathcal{U}^{\dagger}, \label{eq:2c}
\end{align}
\end{subequations}
within $\epsilon$ of the identity using $2d^2$ steps in the sequence of (\ref{eq:1}). The parameter $\Delta$ need not be small and can be less than or equal to $\pi$. The above procedure has been illustrated in Figure \ref{fig2}. We now state a conjecture regarding the linear independence of the generators in $su(d)$ which can be constructed by the non-infinitesimal technique.
\bigskip \\
{\bf Conjecture I.} Suppose $A$ and $B$ are random traceless Hermitian matrices in $\mathcal{H}_d$ and $k \leq N$. Then the set of elements $\{ U^{\dagger} \partial U/\partial t_k, U^{\dagger} \partial U/\partial \tau_k \}_{\vec{t^{\prime}},\vec{\tau^{\prime}}}$ are linearly independent and forms a basis in $su(d)$ for almost any $U(\vec{t^{\prime}},\vec{\tau^{\prime}})$. Here $U = e^{-i B \tau_N}e^{-i A t_N} \ldots e^{-i B \tau_1}e^{-i A t_1}$ and $N = d^2/2$.
\bigskip \\
\textbf{Completeness of a set of a Hamiltonians} \\
{\bf Definition.} A set of Hamiltonians $\{H_i\}$ is complete if there exists a spanning set of nested commutators of $\{H_i\}$ in the corresponding $u(d)$ algebra.
\bigskip \\

We have numerically verified the above conjecture for random matrices $A$ and $B$ in $d = 2,3,4,5$. Note that this method allows us to move a distance $\epsilon$ towards the desired unitary $U$ via gradient descent. The above conjecture also implies the following. For random matrices $A, B$ and $k \leq d^2/2$, the set of all partial derivatives $\{ \partial U/\partial t_k, \partial U/\partial \tau_k \}_{\vec{t^{\prime}},\vec{\tau^{\prime}}}$ spans the space of directions in the manifold of unitaries at that point. Here,
\begin{equation} \label{eq:3}
\frac{\partial U}{\partial t_k} = e^{-i B \tau_N} e^{-i A t_N} \ldots (-i A)e^{-i A t_k} \ldots e^{-i B \tau_1} e^{-i A t_1}.
\end{equation}
Similarly for $\partial U/ \partial t_k$. Thus, we can explore any direction in the space of unitaries in the neighbourhood of $U(\vec{t},\vec{\tau})$ and hence in the neighbourhood of the identity after mapping back by exploiting the first order variation of $U(\vec{t},\vec{\tau})$ with respect to $\vec{t},\vec{\tau}$. If we are given an explicit representation of $U$, this technique allows us to move in the right direction towards the target $U$. However, because the gradients of $U(\vec{t},\vec{\tau})$ form a spanning set, we can also proceed along the right direction using gradient descent on the weight space $\vec{t},\vec{\tau}$ to reduce the error function defined by the training set. This will be explained in more details below. 

Motivated by the above conjecture, we ask how many linearly independent directions in the algebra of Hamiltonians in $\mathcal{H}_d$ can be explored when the matrices $A$ and $B$ correspond to nearest neighbour random local interactions. Specifically, for a $n$-qubit physical system on a 1-dimensional lattice, we consider the following pair of Hamiltonians,
\begin{subequations} \label{eq:2}
\begin{align} 
A_{loc} &= A_{1,2} \otimes \mathcal{I}_{3 \cdots n} + \mathcal{I}_{12} \otimes A_{3,4} \otimes \mathcal{I}_{5 \cdots n} + \cdots + \mathcal{I}_{1 \cdots n-2} \otimes A_{n-1,n} ,
\\
B_{loc} &= \mathcal{I}_{1} \otimes B_{2,3} \otimes \mathcal{I}_{4 \cdots n} + \mathcal{I}_{123} \otimes B_{4,5} \otimes \mathcal{I}_{6...n} + \cdots + \mathcal{I}_{1 \cdots n-3} \otimes B_{n-2,n-1} \otimes \mathcal{I}_{n} ,
\end{align}
\end{subequations}
where  $A_{i,i+1}$, $B_{j,j+1}$ are two-qubit random traceless Hamiltonians and $\mathcal{I}$ is the identity matrix. Similar to the previous case, random matrices of the form $A_{loc}$ and $B_{loc}$ enables one to move along all $d^2$ linearly independent directions in the unitary manifold. 
 \bigskip \\
{\bf Conjecture II.}
Suppose $A_{loc}$ and $B_{loc}$ are random traceless Hermitian matrices in $\mathcal{H}_d$ and $k \leq N$. Then the set of elements $\{ U^{\dagger} \partial U/\partial t_k, U^{\dagger} \partial U/\partial \tau_k \}_{\vec{t^{\prime}},\vec{\tau^{\prime}}}$ are linearly independent and forms a basis in $su(d)$ for almost any $U(\vec{t^{\prime}},\vec{\tau^{\prime}})$. 
 \bigskip \\
We have numerically verified the above for $n$ = 3,4 qubits. We expect Conjecture II to hold true for $n > 4$. We comment that Conjecture II remains valid even when $A_{loc}$ and $B_{loc}$ comprises of local homogeneous Hamiltonians.  By homogeneity, we mean that all $A_{i,i+1}$ and all $B_{j,j+1}$ are one and the same two-qubit random local transformation. 

 An interesting feature of the expression $\mathcal{C}$ in (\ref{eq:2}) is its resemblance with a physical quantity called the out-of-time-ordered correlator (OTOC). For Hermitian or unitary operators $V$ and $W$, the OTOC is defined as the the correlation function $F_t = \langle W_t^\dagger V^\dagger W_t V \rangle = Tr(W_t^\dagger V^\dagger W_t V \rho)$ where $\rho$ is the state of the physical system, $W_t = U^{\dagger} W U$ and $U$ is the time evolution operator. OTOCs are used to characterize the delocalization or scrambling of quantum information in strongly interacting many-body quantum systems via the exponential growth of local Heisenberg operators. The ability to attain any direction in $su(d)$ is equivalent to the ability in using our controls to perform scrambling or effective randomization of the many-body dynamics.

\subsection{Learning U from training data}
Suppose, instead of an explicit representation of $U$, we have access to example input and output pairs, $|\psi_\ell\rangle$, $U|\psi_\ell\rangle$ as a training set. In such cases, gradient descent can be performed using the error function $ E = 1- (1/M) \sum_\ell \langle \psi_\ell| U^\dagger U(\vec t, \vec \tau) |\psi_\ell\rangle$, where $M$ is the number of elements of the training set. For the general criteria of when such a training set allows one to learn about $U$ exactly, see \cite{Marvian}.  As long as $N = O(d^2)$, we can explore the full neighborhood of the identity to reduce $E$. 

Our aim is to minimize $E$ by moving along the right direction towards $U$ such that $||U | \psi \rangle - e^{-i H \epsilon} | \psi \rangle|| < 1$. The first step is to randomly assign parameters $\{\vec{t}, \vec{\tau} \}$ in the non-infinitesimal sequence and perform gradient descent in the error function $E$. This method allows us to move a distance $O(\epsilon)$ closer to $U$. There can be situations when for a given initial parameter choice, the gradient descent method may not converge to the target unitary $U$ because of the existence of saddle points in the weight space ${\vec{t}, \vec{\tau} }$ of $E$. Having advanced as far as we can in the direction of $U$ via the first sequence, we add another non-infinitesimal sequence of fixed depth $O(d^2)$ by assigning different parameters and perform gradient descent again. Since we can explore any direction in the vicinity of identity using the non-infinitesimal method, this ensures that we can learn about U in $O(d^2/\epsilon)$ steps by gradient descent method.


\section{Construction: Infinitesimal Case}
The non-infinitesimal method elucidated in the previous section scales optimally as $N = O(d^2)$. Traditional methods for building unitaries rely on infinitesimal methods using nested commutators \cite{Lloyd3,Sefi,Childs1}. Moreover, as noted above, the infinitesimal approach allows us to address the problem of power and band width limits via the process of Trotterization. Here we demonstrate that such methods scale only slightly worse than optimal. The number of steps required in (\ref{eq:1}) to realize a unitary near identity scales as $N = O(d^2 \log d)$. 

The basic method for constructing unitary transformations close to the identity relies on the Campbell-Baker-Hausdorff relation
\begin{equation} \label{eq:4}
e^{i B t}e^{i A t}e^{-i B t}e^{-i A t} = e^{[A,B]t^2} + O(t^3).
\end{equation}
Repeated application of this method allows the implementation of effective Hamiltonians which take the form of nested commutators, for example
\begin{equation} \label{eq:5}
[A, [A, [B, [A, [B, [ \ldots [A,B]] \ldots ],
\end{equation}
to accuracy $t^k$ where $k$ is the number of commutators in the above $k+1$-th degree homogeneous polynomial. The number of operations required in the sequence of (\ref{eq:1}) to implement a $k$-th degree nested commutator is $N = O(2^k)$. There are a total of $O(2^k)$ nested commutators till order $k$. Thus when $2^k = O(d^2)$ such that $k = 2 \log d + c$ where $c = O(1)$, there are potentially $d^2-1$ nested commutators to span all directions in the $su(d)$ algebra in the neighbourhood of the identity. In the rest of the paper, the base of logarithm is 2. The desired commutator appears with a coefficient $t^k$ and error terms appear from $t^{k+1}$. 

Notice that not all of the above nested commutators are linearly independent. For example, $[A,[B,[A,B]]] = -[B,[A,[A,B]]]$. We found that there are two sources of redundancies in the $k$-th degree nested commutators. First, the redundancies that are due to nested commutators of lower degrees. Second, there can be new internal redundancies as well. An example of a new internal redundancy at sixth order is $ [B[A[A[B[A, B]]]]] = - [A[B[A[B[B, A]]]]] + \frac{1}{3} \Big([A[A[B[B[B, A]]]]]  +  [B[B[A[A[A, B]]]]] \Big)$, which is independent of redundancies at lower orders. Further examples of higher order linearly independent nested commutators have been provided in Appendix A. 

To count the number of such linear dependencies amongst the nested commutators at order $k$, we use a theorem due to Witt \cite{Witt}.
\bigskip \\
{\bf Theorem.}(Witt) Suppose $L(X)$ is a free Lie algebra on a $q$-element set $X$. $a_k$ is the dimension of the homogeneous part of degree $k$ of $L(X)$. Then $a_k = \frac{1}{k} \sum_{ \mathfrak{d}/k } \mu(k/ \mathfrak{d})(q^{\mathfrak{d}} - 1)$. 
\bigskip 

The important point here is that $a_k$ is the number of potentially linearly independent nested commutators of order $k$. Here $\mu$ is the M\"obius function, $q$ is the cardinal number of the generating set $X$, $\mathfrak{d}$ is the $m$-th divisor of $k$ and the summation is over all divisors $\mathfrak{d}$ of $k$.  Further details about a free Lie algebra have been presented in Appendix B. A standard reference for free Lie algebras is \cite{Reutenauer,oeis}. Although the expression in Witt's theorem is rather complicated, it immediately implies the following.
\bigskip \\
{\bf Corollary.} The cardinality of the spanning set of $k$-th degree nested commutators of a free Lie algebra $L(X)$ scales as $O(q^k)$. 
\bigskip 

Consequently, the number of potentially linearly independent nested commutators scales as $O(d^2)$. We conjecture that for randomly selected traceless Hermitian matrices $A$ and $B$ in $su(d)$, the nested commutators till degree $k = O(\log d)$ which forms the basis elements of the free Lie algebra $L$ are actually linearly independent in the Lie algebra $su(d)$. Thus for random $A$ and $B$ operators, we can obtain $d^2-1$ basis elements of $su(d)$ from the spanning set of the first $k$-th degree nested commutators where $k = 2 \log d + c$. That is, we make the following statement.
\bigskip \\
{\bf Conjecture III.} Suppose $A$ and $B$ are random traceless Hermitian matrices in $\mathcal{H}_d$. Then the set of all linearly independent nested commutators in the free Lie algebra of $A$ and $B$ are linearly independent in $su(d)$. 
\bigskip 

We have verified this conjecture numerically for $d$ up to 20 and order $k$ up to 11. The number of linearly independent nested commutators can be further reduced if the operators $A$ and $B$ have a specific form, for example, when $A$ and $B$ are small order polynomials of the Pauli operators. In all cases that we have investigated, however, the number of linearly independent nested commutators of degree $k$ scales as $O(2^k)$. Additional lack of linear independence simply increases the $O(1)$ constant $c$ above. We have not come across an example of a set of matrices $\{X_a\}$ that generates the full Lie algebra by commutation but that fails to generate a linearly independent spanning set of operators via nested commutators of order $k = O(\log d)$.

 The constructive infinitesimal procedure for implementing a desired unitary is as follows. We choose the order $k$ with $2^k = O(d^2)$ such that a spanning set for the Lie algebra $su(d)$ is generated by commutators within order $k$ of the sequence in (\ref{eq:1}). It takes $N=O(d^2)$ transformations to enact such commutators. The commutators at order $k$ occur with coefficients that scales as $O(t^k)$. The higher order terms beyond order $k$ generate transformations that are spanned by the lower order terms. That is, using the infinitesimal method, we can generate some particular unitaries in the vicinity of the identity.   Accordingly, there exists an $\epsilon$ ball in the neighbourhood of identity where a specific set of unitaries within the ball can be implemented using $N=O(d^2)$ tranformations. 
 
Note that the obtainable $\epsilon$-ball is larger than $t^k$ but the constructive method only allows us to produce some particular unitary transformations explicitly with coefficient $t^k$. The size of the actual $\epsilon$-ball reachable at order $k = O(\log d)$ is given by $\epsilon_f = \epsilon^k$. We rescale the size of the $\epsilon$-ball by $\tilde{\epsilon} = |\log \epsilon|$ such that $\tilde{\epsilon}$ is small. This gives the following relation $\tilde{\epsilon} = \tilde{\epsilon}_f / \log d$ which implies that a minimum time of $O(d^2 \log d/ \tilde{\epsilon}_f)$ is required to implement a finite unitary transformation. Thus we are able to build a particular subset of all possible unitaries in $SU(d)$. It would be interesting to quantify how does the measure of the space of unitaries generated  via the infinitesimal approach in time $O(d^2 \log d/ \tilde{\epsilon}_f )$ scales with the dimension of $\mathcal{H}_d$.

\section{Applications}

\begin{figure}[t] 
\includegraphics[width=0.45\linewidth,height=5cm]{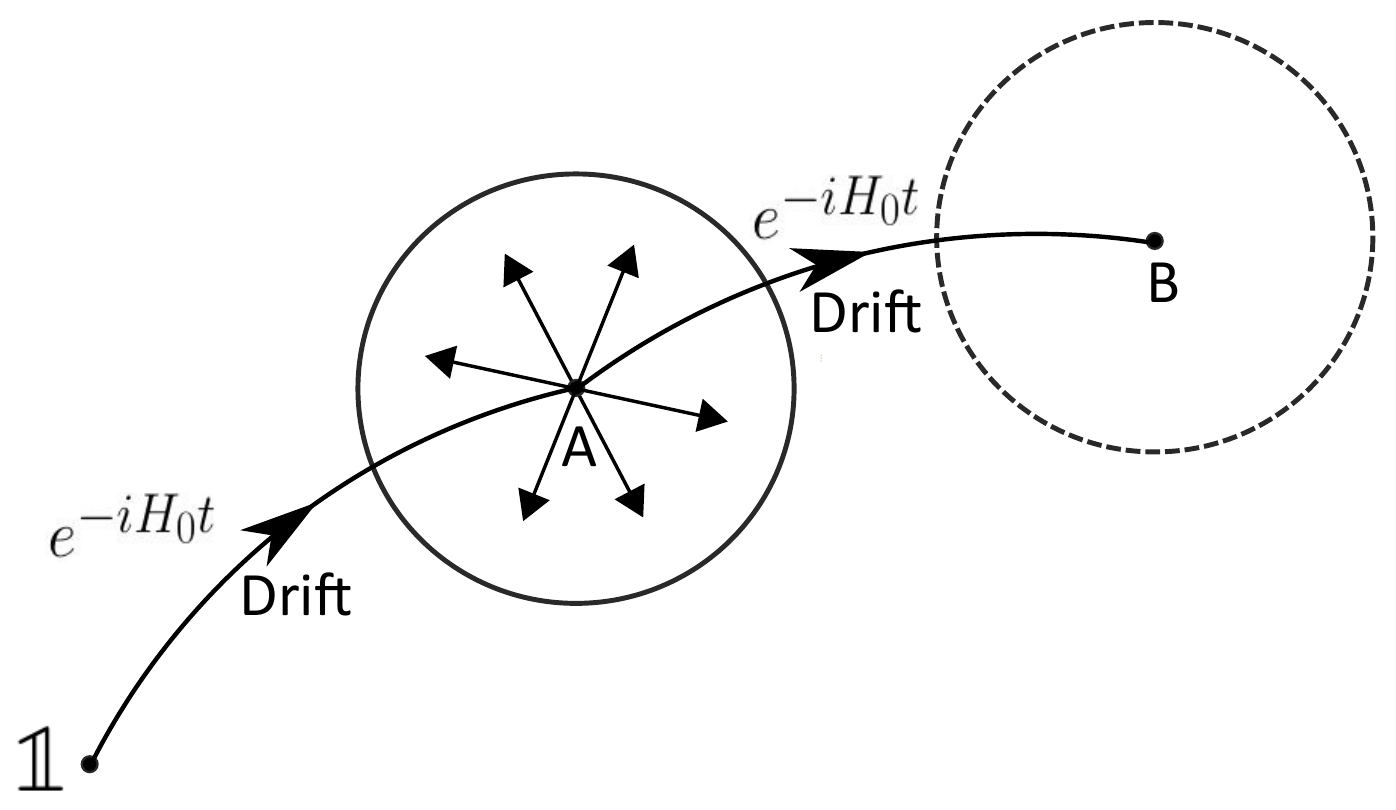}  
\caption{Unitary evolutions that can be realized in the presence of a drift Hamiltonian $H_0$.}
\label{fig3}
\end{figure}
In quantum control, one considers a time-dependent Hamiltonian of the form $H = H_0 + \gamma(t) H_c$ where $H_0$ and $H_c$ are the drift and control Hamiltonians, and $\gamma(t)$ is a time-dependent control field \cite{Lloyd4}. The objective is to implement unitaries in $SU(d)$ via the non-infinitesimal method using bounded Hamiltonians of the form $H_0 + \gamma(t) H_{a}$ and $H_0 + \gamma(t) H_{b}$. We make the following assumptions. First, the inverse transformation of the drift Hamiltonian $H_0$ cannot be implemented. Second, $[H_0,H_{a}] \ne 0$, $[H_0,H_{b}] \ne 0$ and $[H_{a},H_{b}] \ne 0$. Third, the control field parameter $\gamma(t)$ is fixed when the parameters $\{t_i,\tau_j \}$ are small. 

We apply unitary transformations corresponding to the Hamiltonians $H_0 + \gamma(t) H_{a}$ and $H_0 + \gamma(t) H_{b}$ according to (\ref{eq:2a}). This technique generates $d^2-1$ directions in the first order such as,
\begin{equation} \label{eq:6}
e^{i (H_0+\gamma_0H_{a})t_1}e^{i (H_0+\gamma_0H_{b})\tau_1} \ldots e^{i (H_0+\gamma_0H_{a})t_{N}} H_{b} e^{-i (H_0+\gamma_0H_{a})t_{N}} \ldots e^{-i (H_0+\gamma_0H_{b})\tau_1}e^{-i (H_0+\gamma_0H_{a})t_1}.
\end{equation}
where, for simplicity, we have set $\gamma(t) = \gamma_0$ for small times. The parameters $\{t_i,\tau_j \}$ have to be small for bounding the error terms such as $[H_0,H_{a}]$, $[H_0,H_{b}]$ and their higher order counterparts. Since we can explore a restricted parameter space in $\vec{t},\vec{\tau}$, we can construct only a specific set of unitaries in the vicinity of the identity in $O(d^2)$ steps. It would be interesting to be able to classify those unitaries that are reachable in time of $O(d^2)$ simulated by Hamiltonians with a drift term $H_0$. If the assumptions $[H_0,H_{a}] \ne 0$ and $[H_0,H_{b}] \ne 0$ are relaxed such that the drift term $H_0$ commutes with the control fields $H_a$ and $H_b$, then we can explore any direction in the manifold of SU(d) in the vicinity of the point $A$ as illustrated in Figure \ref{fig3}. Note that when $\vec{t},\vec{\tau}$ are not small, we need to implement the time ordered exponential
\begin{equation}  \label{eq:7}
\mathcal{T} \mbox{exp} \Bigg\{ i \int_{t=0}^{t=T} (H_0 + \gamma(t) H_a)dt \Bigg\} = \prod_{i=1}^{m} \mbox{exp} \Bigg\{i \Big(H_0 + \gamma_i H_a \Big)\Delta t \Bigg\},
\end{equation}
where $\mathcal{T}$ is the time-ordering operator and $T = m \Delta t$. Similarly for $H_0 + \gamma(t) H_b$. The number of steps now would scale as $O(m d^2)$ to construct any unitary near the identity. Since the $SU(d)$ group is compact, any desired unitary can be constructed with the non-infinitesimal approach.

A recent study by Zhu {\it et al.} \cite{Zhu} showed how to measure out-of-time-ordered correlators (OTOCs) by appending an ancilla qubit to the physical system of interest. The ancillary state controls the overall sign of the Hamiltonian of the physical system by functioning as a `quantum clock'. The authors indicated how to measure the correlator $\langle e^{i H t} O_2 e^{-i H t} O_1 e^{i H t} O_2 e^{-i H t} O_1 \rangle$ for operators $O_1$ and $O_2$ in the Ramsey interferometry protocol. They demonstrated experimental realization of their protocol in cavity-QED systems such as local XY spin chains or extended Bose-Hubbard model. It would be interesting to implement their protocol for reversing the sign of the physical Hamiltonian in the non-infinitesimal approach for generating arbitrary unitary transformations. 

\section{Conclusion}
The ability to simulate arbitrary unitary operations using polynomial resources in time can be a stepping stone towards developing a NISQ computer \cite{Preskill}. In this work, we have demonstrated two approaches based on sequential application of Hamiltonian operations for realizing unitary transformations in $SU(d)$. The evolution times of the Hamiltonians are systematically controlled in order to reach the desired unitary accurately in $O(d^2/ \epsilon)$ steps for the non-infinitesimal method while the scaling is slightly worse for the infinitesimal approach. Recall that the Solovay-Kitaev theorem shows how to approximate an arbitrary transformation in $SU(d)$ to accuracy $\delta$ with $O(d^2 \log^{3.97}(1/ \delta)$ elementary one and two qubit gates \cite{Dawson}. We have shown that the non-infinitesimal technique achieves the same scaling with dimension $d$ as the Solovay-Kitaev algorithm to construct any desired unitary. Note that $\epsilon$ and $\delta$ are not related to each other.

The primary difference between the infinitesimal and non-infinitesimal procedures is that the infinitesimal method requires one to construct nested commutators at order $k$ of the sequence in (\ref{eq:1}). The $k$-th order nested commutators occur with small coefficients $\delta^k$ where the ratio between the desired term and higher order error terms is $O(\delta)$. In contrast, the non-infinitesimal approach generates all $d^2-1$ basis elements in the first order. This implies we can follow any direction in the space of unitaries by varying the parameters $\{t_i,\tau_j\}$ to the first order. A drawback of the infinitesimal approach is one must explore high-order saddle points of a complex landscape to achieve a desired unitary transformation via gradient descent method. The convergence of gradient descent in such a landscape is difficult to prove \cite{Anandkumar}. In contrast, the non-infinitesimal method provides a direct way for finding optimal solutions in the vicinity of identity via gradient descent. Another notable point is that in the ideal case, the non-infinitesimal method can generate any unitary exactly while the infinitesimal method is associated with an approximation error. 

Both the infinitesimal and non-infinitesimal methods can generate directions in the $SU(d)$ manifold that are either nested commutators or their linear combinations which is a Lie polynomial. An intriguing question one can consider in this direction goes as follows, how hard is it to generate the exponential of the operator $A^m B^n \pm B^n A^m$ where $A$ and $B$ are bounded hermitian matrices. This is intrinsically related to the question of Hamiltonian complexity for simulating time-evolution of physical systems whose dynamics is governed by the operator $A^m B^n \pm B^n A^m$. Another future direction related to the current work is to analyse the sensitivity of our approaches in the presence of experimental noise. This is an important question since we can only implement the Hamiltonians $A, B$ and their time evolutions $\vec{t},\vec{\tau}$ with a finite precision in experiments. The above questions are left open and will be considered for future studies.
 
\section{Acknowledgements}
RM would like to thank Mark Wilde, Zi-Wen Liu and Vlatko Vedral for insightful conversations. The work of RM was supported by a Felix Scholarship and a Great Eastern Scholarship from the University of Oxford.


\begin{appendix}
\section{PLI Nested Commutators}
In this appendix, we have tabulated the linearly independent nested commutators for random matrices $A$ and $B$ in $su(d)$.

\begin{table}[h]
\centering
    \begin{tabular}{ | c | C{7cm} | C{7cm} | C{1.5cm} | N }
    \hline
    Order  & Commutators & Linearly Independent & $N_{min}$ &\\[10pt] 
    \hline
    1 & A, B & A, B &  &\\[15pt] 
    \hline
    2 & [A,B] & [A,B] &  4  &\\[15pt]
    \hline
    3 & [A,[A,B]], [B,[A,B]] & [A,[A,B]], [B,[A,B]] &  8 &\\[15pt]
    \hline
    4 & [A,[A,[A,B]]], [B,[B,[A,B]]], [A,[B,[A,B]]], [B,[A,[A,B]]] & [A,[A,[A,B]]], [B,[B,[A,B]]], [A,[B,[A,B]]] &  12 &\\[20pt]  
    \hline
    5 & [A,[A,[A,[A,B]]]], [B,[B,[B,[A,B]]]],  [A,[B,[A,[A,B]]]], [B,[A,[B,[A,B]]]],  [A,[A,[B,[A,B]]]], [B,[B,[A,[A,B]]]], [B,[A,[A,[A,B]]]], [A,[B,[B,[A,B]]]], [[A,B],[A,[A,B]]], [[A,B],[B,[A,B]]] & [A,[A,[A,[A,B]]]], [B,[B,[B,[A,B]]]], [A,[A,[B,[A,B]]]], [B,[B,[A,[A,B]]]], [B,[A,[A,[A,B]]]], [A,[B,[B,[A,B]]]] &  $\sim$ 18 &\\[50pt]  
    \hline
    6 & [A,[A,[A,[A,[A,B]]..], [B,[B,[B,[B,[A,B]]..], [[A,[A,B]],[B,[A,B]]], [A,[[A,B],[A,[A,B]]],... & [A,[A,[A,[A,[A,B]]..], [B,[B,[B,[B,[A,B]]..], [A,[B,[B,[B,[A,B]]..], [A,[A,[B,[A,[A,B]]..], [A,[B,[A,[B,[A,B]]..], [A,[B,[A,[A,[A,B]]..], [A,[A,[B,[B,[A,B]]..], [B,[B,[A,[B,[A,B]]..], [B,[B,[A,[A,[A,B]]..] &  $\sim$ 27 &\\[50pt]  
    \hline
     \end{tabular}
\caption{The first column represents the order of the Taylor expansion of Eqn.(\ref{eq:1}), the second column includes all possible commutators for a given order $k$, the third column indicates the linearly independent commutators and the fourth column specifies the minimum number of parameters $N_{min}$ required to exponentiate a $k$-th degree nested commutator.}
     \label{table:1}
\end{table}

\section{Free Lie Algebra}
The contents of this appendix is standard literature \cite{Gomis,Reutenauer,oeis,wiki,Berstel}. For the sake of completeness, we define a free Lie algebra that has been previously studied in the literature in the context of out-of-time-ordered correlators (OTOCs) \cite{Haehl} and in relation to Maxwell algebra \cite{Gomis}.  

A free Lie algebra $L(X)$ is the maximal Lie algebra that can be constructed over a generating set $\{X_a\}$ where $a=1,2,..,q$ such that skew symmetry and Jacobi identity holds. Since the generators do not satisfy any additional imposed relations, the free Lie algebra is infinite dimensional and it is the linear space spanned by nested commutators of the form $[X_{a_i},[X_{a_j},[\ldots[X_{a_l},X_{a_m}]]\ldots]$. The homogeneous part of degree $k$ of the free Lie algebra refers to a free Lie subalgebra that is spanned by $k$-th degree nested commutators. For example, the free Lie subalgebra spanned by all commutators of the form $[X_a,X_b]$ is a space of dimension $q(q-1)/2$.
\bigskip \\
{\bf Definition.} A free Lie algebra $L$ on a set $X$ is a Lie algebra with a mapping $i : X \to L$ that satisfies the following universal property. For every Lie algebra $M$ with the mapping $g : X \to M$, there exists a unique Lie algebra homomorphism $G : L \to M$ such that $g = G \circ i$. 
\bigskip 

A general property of the map $G$ when $|X|$ = dim($M$), $G$ is surjective. One can prove that there exists a unique free Lie algebra $L$ generated by a set $X$. The basis elements of a free Lie algebra $L$ can be constructed in terms of Lyndon words which gives the Lyndon basis. Lyndon words have wide ranging applications in algebra and combinatorics.
\bigskip \\
{\bf Definition.} A Lyndon word is a primitive word which is strictly smaller than all its non trivial cyclic rotations.
\bigskip \\
For example, the Lyndon words for the two-symbol binary alphabet $\{a,b\}$ sorted by length and then lexicographically forms an infinite sequence given by,
\begin{equation}
\{a,b\}, \{ab\}, \{aab,abb\}, \{aaab, aabb, abbb\}, \{aaaab, aaabb, aabab, aabbb, ababb, abbbb\}, \ldots
\end{equation}

The number of Lyndon words of length $k$ on $q$ symbols is given by the Witt formula. A Lyndon word $u$ that is not a letter has the following property of being expressed as $u=v w$ where $v,w$ are Lyndon words with $v<w$ lexicographically. In general, this is not a unique factorisation since, for example, $(a)(abb) = (aab)(b)$. However there is a unique factorisation of a Lyndon word $u$ as a product of two Lyndon words $v,w$ with $v<w$ termed as the standard factorisation. An important theorem in this context is given below.
\bigskip \\
{\bf Theorem.}(Chen-Fox-Lyndon) If $w$ is lexicographically the smallest proper suffix of a Lyndon word $u=v w$, then $v$ and $w$ are also Lyndon words such that $v < w$.
\bigskip \\
The standard factorisation of a Lyndon word is obtained by selecting $w$ to be the lexicographically least proper suffix of $u$ which is also the longest proper suffix of $u$ that is a Lyndon word. In the above example,  $(a)(abb)$ is the standard factorisation of $u=aabb$.

There exists a bijection $\mathcal{M}$ from the set of Lyndon words to the basis elements of a free Lie algebra. $\mathcal{M}$ is defined as follows. If the word $u$ is a letter, then $\mathcal{M}(u)=u$. When the length of $u$ is greater than or equal to two, then by standard factorisation, $u = v w$ for Lyndon words $v,w$ and $w$ being the longest possible suffix. Thus $\mathcal{M}(u) = [\mathcal{M}(v),\mathcal{M}(w)]$. For example, the standard factorisation $(a)(ababb)$ of the Lyndon word $aababb$ can be mapped to the commutator $[a,[[a,b],[[a,b],b]]]$.

It is evident that the basis elements of free Lie algebra $L$ are homogeneous polynomials in the elements of the generating set $X$. The free Lie algebra vector space can be expressed as a direct sum of graded Lie algebras given by,
\begin{equation}
L = \bigoplus\limits_{k>0}{} L_k = L_1 \oplus L_2 \oplus \ldots 
\end{equation}
where $L_1$ is a $q$-dimensional vector space spanned by the elements of $X$ with $|X|$=$q$. The vector space $L_2$ is spanned by commutators of the form $[X_a,X_b]$ with the corresponding dimension equals $q(q-1)/2$. $L_2$ can also be expressed as $[L_1,L_1]$. Technically, $L_2$ is referred to as the exterior square of $L_1$. Notice that this holds true for any arbitrary $k > 0$,
\begin{equation}
L_{k+1} = [ L_k , L_1 ] .
\end{equation}
By the definition of graded Lie algebras, we have the following relation $[L_k, L_{k^'}] \subseteq L_{k+k^'}$. An important property of any subalgebra of a free Lie algebra is due to Shirsov and Witt.
\bigskip \\
{\bf Theorem.}(Shirshov–Witt) Any Lie subalgebra of a free Lie algebra is a free Lie algebra.
\bigskip \\ 
This is an analogue of the Nielsen-Schreier theorem in group theory which states that every subgroup of a free group is free.
\end{appendix}

\end{document}